\documentclass[twocolumn,superscriptaddress,floatfix,preprintnumbers,amssymb ,amsmath]{revtex4}
\usepackage{graphicx}
\usepackage{dcolumn}
\usepackage{bm}
\usepackage[latin1]{inputenc}
\usepackage[mathscr]{eucal}
\usepackage{epsfig}

\begin{document}

\title{Superconducting transition detector in power amplification mode: a tool for cryogenic multiplexing}

\author{P. Helist\"o}
\affiliation{VTT Technical Research Centre of Finland, PO BOX 1000, 02044 VTT, Finland}
\author{J. Hassel}
\affiliation{VTT Technical Research Centre of Finland, PO BOX 1000, 02044 VTT, Finland}
\author{A. Luukanen}
\affiliation{MilliLab, PO BOX 1000, 02044 VTT, Finland}
\author{H. Sepp\"a}
\affiliation{VTT Technical Research Centre of Finland, PO BOX 1000, 02044 VTT, Finland}

\begin{abstract}
We demonstrate that substantial power gain 
can be obtained with superconducting transition detectors. 
We describe the properties of the detector as a power amplifier theoretically. 
In our first experiments power gain of 23 was reached in a good agreement with the theory. 
The gain facilitates noise matching of the readout circuit to the detectors in the 
case of time division multiplexing.
\end{abstract}


\maketitle


Cryogenic multiplexing is the only practical route towards applications of large format 
superconducting detector arrays (see, e.g., a review of SQUID multiplexers \cite{irwin02}).
Time-division multiplexing (TDM) is one 
of the basic alternatives. In TDM the outputs from $N$ detectors are multiplexed to a single output
cable with the help of a cryogenic switch, scanning sequentially through the detectors.
Thus each detector is only observed for $\sim 1/N$ of the total measurement time. 
As a consequence, the switching operation multiplies the noise temperature of the 
post-switch amplifier by $N$, referred to the noise temperature 
of the pre-switched signals.
This makes noise matching of the readout to the low-noise cryodetectors considerably
more tedious than without 
multiplexing. 

To keep the readout noise contribution well below that of the detectors, additional 
power amplification is needed before the switching operation.
The gain is normally provided by the readout amplifier, but amplification by the 
detector is also possible. In this paper, we characterise the achievable power 
gain, dynamic range, noise and stability of resistively biased transition detectors.

Superconducting transition detectors such as transition edge calorimeters and bolometers 
\cite{mather82,richards94} or superconducting wire bolometers \cite{luukanen03} are usually 
operated in voltage biased mode at voltages much below the minimum of the detector $I - V$ 
curve.\cite{irwin95}
Such biasing provides high current responsivity, low effective detector Johnson-Nyqvist noise, 
high stability 
and good linearity due to the strong negative electrothermal feedback (ETF). However, the 
detector provides no power gain
and the dynamic resistance of the detector is very low (and negative). This calls for a special 
readout device -- the SQUID. 

Recently it was demonstrated that the electrothermal feedback of a voltage-biased
superconducting bolometer at the $I-V$ curve minimum can be used for efficient noise
matching of the detector to standard room temperature readout amplifier in feedback 
mode \cite{penttila06,luukanen06,helisto07}. This is because the output noise temperature of
the transition detector diverges at the minimum \cite{helisto07}. 
Unfortunately this scheme is not well suited to 
large scale cryomultiplexing due to its limited power gain bandwidth.

In this paper, we analyze the properties of a resistively biased transition detector, operated 
in the power amplification mode. As an example we use a 'hot-spot' superconducting wire 
bolometer \cite{luukanen03} , but the results 
are applicable to other types of thermal detectors also, such as transition-edge 
sensors \cite{kiviranta02}.

\begin{figure}[h]
\centering
\includegraphics[width=8cm]{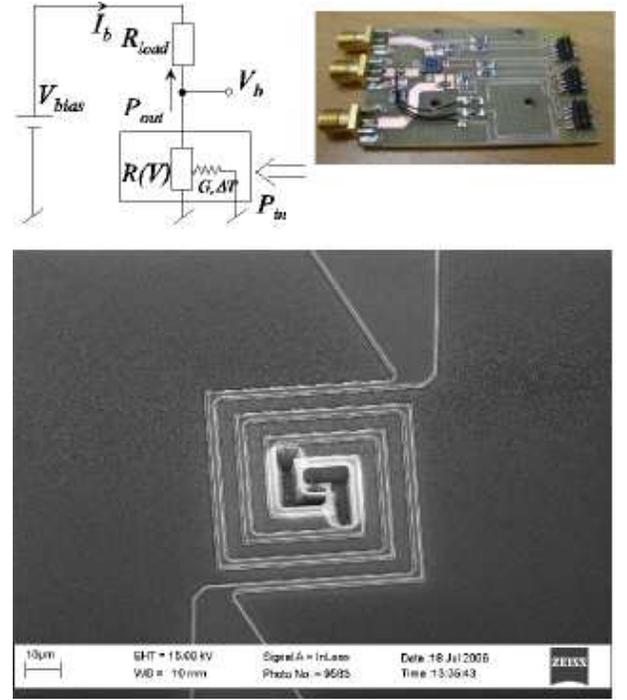}  
\caption{Top left: A schematic picture of the circuit. 
Top right: A photograph of the 2.5 K chip carrier 
containing the chip with three bolometers. 
Bottom: A scanning electron 
micrograph of the NbN airbridge bolometer, coupled to a square-spiral lithographic antenna.} 
\label{expsetup}
\end{figure}

Let us consider the circuit shown in Fig.\ \ref{expsetup} (top left), in which the detector is 
biased through a bias (or load) impedance $R_{load}$ instead of using ideal voltage bias. The detector is
connected to bath temperature $T_b$ with thermal conductance $G$. The critical temperature 
of the superconducting detector is $T_c$ and $\Delta T = T_c - T_b$. The bias point resistance 
of the detector is $R(V)$ and $R_N$ is the normal state resistance of the detector.
Assuming that the detector is a vacuum-bridge microbolometer wire \cite{luukanen03}, its DC
properties can be described by simple electrical and thermal equations
\begin{eqnarray}
v_{bias} &=&\left( r_{l}+r\right) i \label{sahko} \\
i^{2}r+p_{in} &=&1/\left( 1-r\right), \label{termo}
\end{eqnarray}
where $v_{bias} = V_{bias}/V_0$, $i = I_b R_N /V_0$, 
$r_{l}= R_{load}/R_N$, and $r=R(V)/R_N$.
A normal-state region, the length of which is determined by the voltage across the detector,
$v = ir = V_b/V_0$, is formed at the center of the bolometer wire. At very low 
voltages, $v << 1$, the normal phase region shrinks to a hot spot. 
$V_0 = \sqrt{G\Delta T R_N}$ is the voltage across the detector at the $I-V$ curve minimum.
(\ref{termo}) is valid when the transition width is small in comparison to $\Delta T$.

We define the power gain as the change of power dissipated in the 
{\em load} resistor versus the change of the input power $p_{in} = P_{in}/(G\Delta T)$ 
absorbed by the superconducting transition detector. 
When $p_{in} << 1$ the power gain of the detector is
\begin{eqnarray}
G_{P}(\omega) &=& \left|\frac{\partial P_{out}}{\partial P_{in}}\right|_{V_{bias}} = 
2I_b \left| \frac{dI_b}{dP_{in}} \right| R_{load} \notag \\
&=& \left|\frac{-\left(1-\beta\right)L_0}{ 1 +\beta L_{0} +i\omega \tau_{th}} \right|
\label{gainw}
\end{eqnarray}
where $L_0 = 1/v^2$ is the gain of the electrothermal feedback loop and
$\beta = (r - r_{l})/(r + r_{l})$. In (\ref{gainw}), 
$\tau_{th}$ describes the thermal time constant
of the detector, determined by its heat capacity and thermal conductance.

The DC power gain can be written as
\begin{equation}
G_{P,0} = \frac{2/v^2}{1-1/v^2 +1/r_{l}} = 2L_{0}/g_{tot}. \label{gain}
\end{equation}
The parameter $g_{tot} = 1/r_{l} + 1/r_d $ is the total differential conductance of the circuit 
from $V_b$ to ground, and $1/r_d = di/dv = 1 - L_0$. 
Not suprisingly, the power gain diverges at 
\begin{equation}
g_{tot} = 0, \label{instab}
\end{equation}
where the system becomes unstable at DC.
This point corresponds to the voltage $v = 1/\sqrt{1+1/r_{l}}$ over the detector.

The dominant thermal noise mechanisms of the system are well known:
the thermal fluctuation  noise (phonon noise),
Johnson-Nyqvist noise of the detector and Johnson-Nyqvist noise of the bias resistor.
The noise power spectral densities are, as referred to the input, respectively,
\begin{eqnarray}
S_{P,ph} &=& 4G\gamma k_{B}T_{c}^{2} 
\\
S_{P,J} &=& \left( 4G\Delta Tk_{B}T_{c} \right) \frac{ 1+L_{0}}{L_{0}^{3}} \\
S_{P,R_{load}} &=& \left( 4G\Delta Tk_{B}T_{R_{load}} \right) 
\frac{\left(L_{0}^{2}-1\right)^2/L_{0}^{3}}{
 L_{0}+2L_{0}/G_{P,0}-1} 
\end{eqnarray}
assuming that the load resistor is at $T_{R_{load}}$ and that the normal phase
region of the detector wire is at $T_c$ (the temperature distribution 
in the hot spot is neglected). 

Johnson-Nyqvist noise
of the detector can be neglected if $L_0 > 1$. The noise of the load resistance 
is of the same order as the thermal fluctuation noise: 
$S_{P,R_{load}}/S_{P,ph}  \simeq  T_{R_{load}} \Delta T/\gamma T_{c}^{2} \lesssim 1$ 
assuming that $T_{R_{load}} < T_c$.
Here $\gamma$ is a 
factor taking into account the temperature distribution in the heat link (see \cite{mather82}). 

For our intended TDM application, the load resistor would a part of a second thermal circuit
(not shown in Fig. \ref{expsetup}). 
Thus a fourth noise mechanism appears: the thermal fluctuation noise of the second circuit. 
This noise contribution is truly suppressed by the power gain 
$S_{P,ph2} = 4G_2\gamma_2 k_{B}T_{c}^{2}/G_{P,0}^2$, where $G_2$ is the thermal conductance
of the second thermal circuit \footnote{A lower limit on $G_2$ is set by the bias power, 
dissipated in the load resistor:
$G_2/G > i^2 r_l = \left( L_{0}+1\right) /\left( L_{0}+2L_{0}/G_{P,0}-1\right) 
iv$.}.

Power gain reduces the dynamic range of the bolometer compared to the case of strong
negative ETF. An estimate of the linear dynamic range is obtained
by calculating the change in the output power to second order:
$p_{out} \simeq a p_{in} + b p_{in}^2 $ 
and by setting $p_{in,max} =|b/a|$.
In the case of ideal voltage bias this gives
$p_{in,max} = 2\left(L_{0}+1\right)^{2}/\left(L_{0}\left( L_{0}+2\right)\right)$. 
When increasing the power gain, the maximum tolerated input power is reduced by
a factor $\sim G_{P,0}^2/2$. The reduced dynamic range may be a problem 
in the case of calorimeters, with which maximum dynamic range and linearity
is typically required.

The signal bandwidth of the detector is given by (\ref{gainw}). In terms of
the power gain and the loop gain, the effective time constant becomes
$\tau_{\rm eff} = (1+G_{P,0})/(1+L_{0}) \tau_{th}$.
For most applications, such as THz imaging or astrophysics, 
the thermal bandwidth of superconducting transition detectors is sufficient. 

The experimental setup is shown in
Fig. \ref{expsetup}. The detectors were antenna-coupled superconducting
THz bolometers cooled down to $T_b \simeq 2.5$ K with a pulse-tube cryocooler. 
The detector element was a piece of NbN film wire with nominal
dimensions (length x width x thickness) 24 x 2 x 0.22 $\mu$m$^ 3$ and $T_c \simeq 12$ K. 
To decrease the thermal conductance, the wire is released from the Si substrate 
by plasma etching \cite{helisto07}. 

The series connection of the bolometer and 
$R_{load}$ was voltage biased with a small bias resistor $R_{bias} << R_{load}$, not
shown in the simplified circuit diagram of Fig. \ref{expsetup}.
To avoid uncertainties caused by the poorly known THz power coupling, a frequency of 100 MHz 
was chosen for detector excitation. This frequency is well above the thermal cutoff 
frequency of the bolometer. The excitation was coupled 
from the 50 $\Omega$ RF source at room temperature to the low temperature stage through a coaxial 
cable with a transition to microstrip lines on the FR4 chip carrier. Finally, the RF signal was coupled 
to the bolometer via wire bonds.
The source was driving $R_{load}$ and the bolometer in parallel. 
The main limitation of power coupling was due the 
impedance mismatch of the bolometer-load configuration. 

\begin{figure}[h]
\centering
\includegraphics[width=8cm]{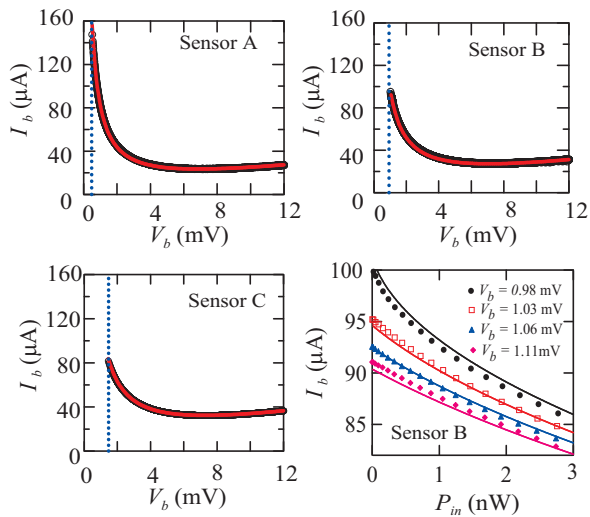}  
\caption{Measured (circles) and fitted (solid curves) DC $I-V$ curves of 
superconducting vacuum-bridge bolometers. 
Fit results and corresponding load resistor values: 
detector A: $R_{load} = 2.83$ $\Omega$, $R_N = 581$ $\Omega$, $G\Delta T = 81$ nW;
detector B: $R_{load} = 9.45$ $\Omega$, $R_N = 519$ $\Omega$, $G\Delta T = 96$ nW;
detector C: $R_{load} = 18.85$ $\Omega$, $R_N = 448$ $\Omega$, $G\Delta T = 118$ nW.
The vertical dotted lines describe the theoretical onset of DC instabiliy according to 
(\ref{instab}).
Bottom right: $I_b$ as a function of applied RF power at different bias points $V_b$ for
detector B. Solid lines are theoretical curves according to (\ref{sahko}) and (\ref{termo}).
}
\label{IVcurve}
\end{figure}

The electrothermal parameters of the bolometers were deduced from $I - V$ curve fits 
to $i = v + 1/v$ (see Fig.\ \ref{IVcurve}). 
Three nominally similar devices (Fig. \ref{expsetup} bottom) on a single chip were 
connected to different load resistances.
Typical results were 
$R_N  \simeq  500$ $\Omega$  and $G \Delta T \simeq 100$ nW.

\begin{figure}[h]
\centering
\includegraphics[width=8cm]{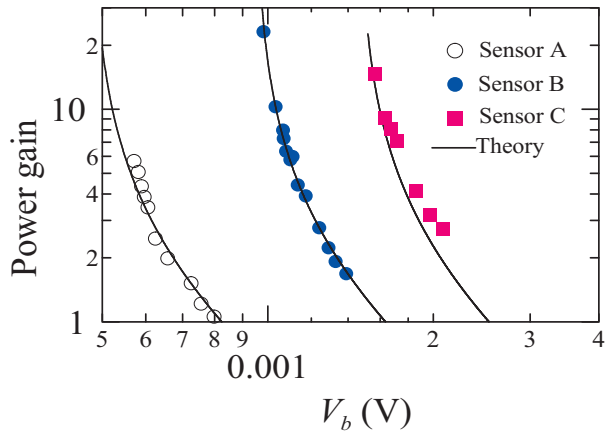}  
\caption{Measured and theoretical power amplification 
as a function of $V_b$ for detectors A, B and C. 
Solid curves are according to (\ref{gain}).}
\label{gains}
\end{figure}

From the experimental 
$I_b$ --  $P_{in}$ curves such as the one shown in Fig. \ref{IVcurve} (bottom right), power gain 
values were obtained as $G_{P,0} =  2R_{load} I_b |\Delta I_b/\Delta P_{in}|$
when $P_{in} \rightarrow 0$. 
The experimental data are in good agreement with the theory (solid curves). 
The curve with detector voltage nearest to the onset of instability 
gives a gain of $G_{P,0}\simeq 23$. 
The resulting gain data as a function of the operating point $V_b$ for the three bolometers 
are shown in Fig. \ref{gains} with the corresponding theoretical curves. Again, a good agreement
is found.

A first look at Eq.\ (\ref{gain}) suggests 
that the loop gain should be maximized to maximise the gain. 
However, with very large $L_0$ and small $R_{load}$
stable biasing becomes more challenging and the gain becomes more sensitive
to small bias voltage variations.
Therefore compromises with the load resistance and 
stability are required. In our experiments, it turned out that
the maximum observed 
power gain depended little on the load resistance value.
When using an antenna-coupled device, the detector resistance should be close 
to the the antenna impedance for best performance.

The bath temperature showed oscillations of about 100 mK due to the pulse tube
cryocooler operating cycle. These oscillations, 
corresponding to $\sim$1 \% of the saturation power of the detectors,
mainly limited our maximum experimental power gain. 
An interesting limit of validity of the model is reached, when the length of the hot spot becomes 
of the order of superconducting tunneling length. 

We have recently proposed a cryomultiplexing 
scheme \cite{luukanen08}, in which the thermal power gain is exploited. 
The load resistor is part of a second stage bolometric amplifier/cryoswitch. 
In this scheme, a power gain of $<50$ is sufficient to 
multiplex at least 100 pixels. One possible application is indoors passive THz imaging for
detection of concealed weapons. The maximum signal change is the difference between
the body and room temperature, about 15 K. Assuming a detection 
bandwidth of $\Delta \nu = 0.5$ THz, detection efficiency of 20 \%, and single mode coupling,
this corresponds to power change of $\Delta P = 0.2 k_B (T_{body} - T_{RT})\Delta \nu
\simeq 20$ pW. Such a power change will not cause a significant nonlinearity at the output of a 
typical 4 K bolometer at power gain values $\lesssim 50$. The main advantages of 
utilizing the thermal power gain for cryomultiplexing 
are process compatibility and low power consumption, when
compared with other types of amplifiers, such as SQUIDs \cite{irwin02} or cryogenic semiconductors
\cite{kiviranta06}.

Since our detectors and the setup were not optimized for the power gain experiments, 
considerably higher power gains should be achievable than are reported here. 
In conclusion, the power amplification mode of superconducting
transition detectors is a feasible tool for a practical, possibly monolithic cryomultiplexer.

We thank Prof.\ Paul Richards and Mr.\ Mikko Kiviranta for useful comments and
Mr.\ Leif Grönberg for detector fabrication.
The work was partly supported by the European Space
Agency, Contract No. 20525/07/NL/CO and
by the Academy of Finland
(Centre of Excellence in Low Temperature Quantum Phenomena and Devices).




\end{document}